# See SIFT in a Rain

Wei Wu, *Member, IEEE*, Hao Chang, and Zhu Li, *Senior Member, IEEE*

*Abstract*—Rain streaks bring complicated pixel intensity changes and additional gradients, greatly obstructing the extraction of image features from background. This causes serious performance degradation in feature-based applications. Thus, it is critical to remove rain streaks from a single rainy image to recover image features. Recently, many excellent image deraining methods have made remarkable progress. However, these human visual system-driven approaches mainly focus on improving image quality with pixel recovery as loss function, and neglect how to enhance image feature recovery ability. To address this issue, we propose a task-driven image deraining algorithm to strengthen image feature supply for subsequent feature-based applications. Due to the extensive use and strong practicability of Scale-Invariant Feature Transform (SIFT), we first propose two separate networks using distinct losses and modules to achieve two goals, respectively. One is difference of Gaussian (DoG) pyramid recovery network (DPRNet) for SIFT detection, and the other gradients of Gaussian images recovery network (GGIRNet) for SIFT description. Second, in the DPRNet we propose an alternative interest point loss that directly penalizes scale response extrema to recover the DoG pyramid. Third, we advance a gradient attention module in the GGIRNet to recover those gradients of Gaussian images. Finally, with the recovered DoG pyramid and gradients, we can regain SIFT key points. This divide-and-conquer scheme to set different objectives for SIFT detection and description leads to good robustness. Compared with state-of-the-art methods, experimental results demonstrate that our proposed algorithm achieves better performance in both the number of recovered SIFT key points and their accuracy.

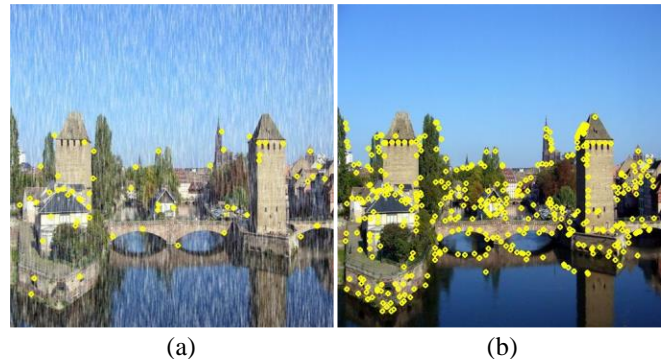

**Fig. 1.** SIFT key points of background extracted from (a) a rainy image and (b) its corresponding clean image.

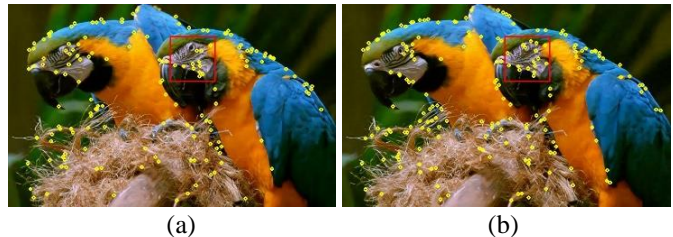

**Fig. 2.** Two derained images obtained via two different image deraining methods, respectively. (a) The red square box area in a derained image has the PSNR and SSIM values of 31.26 dB and 0.937, respectively, as well as the number of recovered SIFT key points of 12. (b) The red square box area in another derained image has the PSNR and SSIM values of 30.73 dB and 0.922, respectively, as well as the number of recovered SIFT key points of 17. Compared with the red square box area in (a), the one in (b) has lower objective and subjective image qualities, but contains much more recovered image features.

## I. INTRODUCTION

RAIN, a common and widespread weather phenomenon, often occurs in most of the world [1]-[3]. As we know, rain always falls from the sky, so that a rain streak layer will have to be added to an original clean image when captured outdoors in such bad weather, producing a rainy image. For their specular highlights, rain streaks bring complicated pixel intensity changes that inevitably mask background information [4]. As a result, additional gradients introduced by rain streaks largely obstruct the extraction of valuable image features. Fig. 1 compares Scale-invariant Feature Transform (SIFT) key points extracted from the backgrounds of a rainy image and its corresponding clean version. From this figure, it can be seen that useful image features collected from a rainy image are far fewer than those from the corresponding clean image, leading to serious performance degradation in image matching and other image feature-based applications. Therefore, it is critical to remove rain streaks from a single rainy image to recover image features.

As everyone knows, two famous image quality assessment indices, including peak signal-to-noise ratio (PSNR) and structural similarity index measure (SSIM) are usually employed to measure a derained image in image deraining. In essence, these two criteria are based on human visual system (HVS). Additionally, derained images are generally used as input to follow-up computer vision (CV) tasks, e.g. image matching, target recognition and tracking, image fusion, 3-D

This work was supported by the National Natural Science Foundation of China, in part by the 111 Project, and in part by High-Performance Computing Platform of Xidian University. *(Corresponding author: Wei Wu.)*

Wei Wu and Hao Chang are with State Key Laboratory of Integrated Services Networks, Xidian University, Xi'an 710071, China (e-mail: wwu@xidian.edu.cn, chang_hao98@163.com).

Zhu Li is with Department of Computer Science & Electrical Engineering, University of Missouri, Kansas City (e-mail: zhu.li@ieee.org).

reconstruction, and change detection, where these CV tasks are mainly based on image features. Fig. 2 illustrates two derained images generated by using different image deraining algorithms, respectively. From this figure, one can see that although the red square box area in a derained image in Fig. 2 (a) has higher PSNR and SSIM values, its recovered SIFT key points is much fewer, indicating that the objective and subjective qualities of an area in a derained image may not directly reflect the recovery of its image features. Hence, to make derained images better applicable to subsequent tasks, the number of recovered image features, a task-driven indicator, is able to be adopted as an important restoration assessment method in image deraining.

So far, many methods have been proposed to try to clear up rain from a single rainy image. Y. Chen *et al*. [5] described a low-rank appearance model to capture spatio-temporally correlated rain streaks. In [6], a single image deraining framework was designed to get rid of rain streaks by learning the context information of an input. In [7], Y. Luo *et al*. sparsely modelled a rain layer with a mutually exclusive learning dictionary, to differentiate rain streaks and background. Y. Wang *et al*. [8] proposed a tensor-based low-rank model, using the similar repeatability pattern of rain streaks to complete rain elimination.

Recently, with the brilliant success of deep learning applied in various language and vision tasks, learning-based algorithms have also been advanced to tackle the problem of image deraining. X. Fu *et al*. [9] developed an end-to-end deep network architecture focusing on high frequency details to obtain a derained image. Based on the first dataset containing rain-density label information, H. Zhang *et al*. [10] proposed a density-aware multi-stream densely connected CNN-based algorithm for joint rain density estimation and deraining. X. Li *et al*. [11] presented a contextual information-based recurrent neural network (RNN) to remove rain from individual images. A progressive recursive network (PReNet) [12] was proposed to serve as a suitable baseline in image deraining research. H. Zhang *et al*. [13] introduced conditional generative adversarial networks (CGAN) to clear up rain streaks. However, all the approaches mentioned above are devoted to improving the objective and subjective qualities of a derained image, and thus neglect how to enhance image feature recovery ability.

To alleviate this issue, in this study we propose a task-driven image deraining algorithm to improve the recovered image features of derained images. SIFT [14], one of the most well-known image feature description methods, has been widely employed in a great number of image processing applications [15]-[19]. Even in the era of deep learning, SIFT has also a strong and solid signal domain interpretation. Indeed, in some big challenges like Google landmark recognition, SIFT-based object re-identification is still showing its advantage as an excellent handcrafted feature solution, compared with learning-based strategies [20]. Moreover, it has been observed that deep features are currently those that perform the best in terms of accuracy, but SIFT-like features still remain highly competitive today in terms of balance between accuracy, storage, efficiency, and hardware-software flexibility [21]. Therefore, due to the extensive use and strong practicability of SIFT, we focus on how to derain a single rainy image to recover SIFT key points as many as possible.

In this paper, we propose a task-driven approach, namely Image Deraining for SIFT Recovery (IDSR). In this proposed algorithm, we first propose a divide-and-conquer strategy using two separate networks, i.e. difference of Gaussian (DoG) pyramid recovery network (DPRNet) as well as gradients of Gaussian images recovery network (GGIRNet), to primarily work on two tasks, respectively. This is inspired by the important idea of SIFT, which is the DoG pyramid and gradient information of Gaussian image are exclusively employed to realize scale and spatial gradient space extrema detection and description, respectively. So the first task is to recover DoG pyramid especially for detecting SIFT key points, whereas the second one is to recover the gradients of derained Gaussian images especially for generating the descriptors of those detected key points. Second, in the DPRNet an alternative interest point (ALP) loss is proposed based on the notable ALP detector in [22]. Besides this novel loss, several channel spatial attention residual blocks (CSARBs) are also adopted to forge a derained image. Third, in the GGIRNet we put forward a gradient attention module (GAM) to exploit important gradient information. Using this GAM with a gradient-wise loss, we construct a channel gradient attention residual block (CGARB) to produce another derained image. Finally, with these two different derained images via the DPRNet and GGIRNet, respectively, we calculate their DoG pyramid and gradients of Gaussian images, respectively, which are further applied to establish recovered SIFT key points.

Our contributions can be summarized as follows:

- To the best of our knowledge, this is the first study to directly recover SIFT from a single rainy image instead of pixel reconstruction for improving image quality. Different from existing HVS-driven image deraining methods which aim to improve objective and subjective image qualities, our proposed IDSR is a task-driven approach developed to strengthen image feature supply for subsequent feature-based vision applications.
- We propose a divide-and-conquer strategy using two separate networks to specially concentrate on DoG pyramid recovery and Gaussian image gradients recovery, respectively.
- We propose an ALP loss function based on a scale space response extrema detection model to accurately locate SIFT key points, which is especially designed for accomplishing the image feature recovery goal.
- We advance a new attention mechanism, dubbed GAM, to generate an attention mask in gradient domain to adaptively select important gradient regions.
- Compared with state-of-the-art (SOTA) methods, our proposed IDSR achieves better performance in both the number of recovered SIFT key points and their accuracy.

This paper is an extension of our prior work [4], where we make further significant improvements: 1) We propose the GAM, a novel attention mechanism, collecting useful information to capture gradient-wise relationships. This GAM is utilized specifically to help the GGIRNet regain derained Gaussian images well. 2) In our current version we also apply a loss function in the gradient domain instead of the combination of the L1 and SSIM losses used in [4], to further accurately establish the gradients of derained Gaussian images. 3) In [4], five parallel network branches are taken to output five corresponding derained Gaussian images, respectively, resulting in lots of parameters. To reduce the number of network parameters, in this work we change them to only one network to create not these Gaussian images but a derained image. 4) Different from our preliminary work in [4], for simplicity, where partial data of Rain1200 and Rain1400 datasets are randomly selected to use, respectively, we further conduct extensive experiments on all the images of each of these two datasets for more rigorous analysis. 5) Besides the two synthetic datasets, we also select a well-known real-world rainy image dataset, i.e. SPA-Data, to evaluate the performance of our proposed algorithm. Experimental results demonstrate that the proposed IDSR recovers more key points than our conference version.

## II. RELATED WORKS

In recent years, there has been an increasing amount literature on image deraining, which can be broadly divided into two categories, i.e. traditional and deep learning-based approaches.

### A. Traditional Image Rain Removal Approaches

Traditional image deraining algorithms generally take advantage of the prior knowledge of rain, to deal with the issue of its elimination from a single rainy image. By appropriately formulating rain removal as a morphological component analysis based image decomposition problem, Y. Fu et al. [23] proposed a single image rain removal framework using bilateral filters. Since the direct use of learned rain and non-rain dictionaries produced unwanted edge artifacts, C. -H. Son et al. [24] developed an image deraining method to shrink sparse codes, generating shrinkage maps and correlation matrices to reduce those artifacts. D. -Y. Chen et al. [25] proposed a monochromatic image-based rain elimination framework, which first decomposed a color image into low-frequency and high-frequency parts, and then learned through both a dictionary and sparse coding to decompose the high frequency part into rain and non-rain components. In [26], several common features of rain and snow were outlined, and a combination of rain and snow detection and a low-pass filter were used to produce low-frequency and high-frequency information, so that those two degradations can be differentiated from the input image. However, as rain has very complicated forms, including various locations, sizes, and orientations, these traditional algorithms obtain rather limited performances.

### B. Deep Learning-based Image Deraining Approaches

Over recent years, the sustainable development of deep learning brings many new powerful solutions to image deraining. Inspired by deep residual networks, X. Fu et al. [9] proposed an end-to-end deep network by changing the mapping form to simplify the learning process, reducing the mapping range between input and output and making the learning process easier. Based on the first dataset containing rain-density label information, in [10] H. Zhang et al. proposed a density-aware multi-stream densely connected CNN-based algorithm, which efficiently leveraged features of different scales for joint rain density estimation and deraining. X. Li et al. [11] presented a recurrent image deraining network based on contextual information for a single rainy image, assigning different alpha values to various rain streak layers by combining squeeze and excitation blocks. In [12] PReNet was proposed to exploit recursive computation and the dependency of depth features across stages by unfolding a shallow residual network. H. Zhang et al. [13] proposed a CGAN-based framework that adopted a densely connected generator to clear up rain, and a multi-scale discriminator to decide whether the corresponding rain removal image was real or false.

In addition, Y. Ye et al. [27] proposed an algorithm that jointly learned rain generation (forward) and rain removal (inverse) in a unified framework. Learning physical degradation can better approximate real rainfall in an implicit manner. In [28], an adaptive dilated network was proposed to remove rain patterns, by constructing an efficient adaptive dilated block, exploiting the importance of different scale features, and modelling the interdependence of adaptive ground features. Y. Yang et al. [29] proposed a progressive residual detail supplementation based end-to-end rain removal network to progressively dislodge rain layers. In [30], a residual multi-scale image deraining method was proposed, in which the residual between the reconstructed image and the input rainy image was treated as an attention map, providing help in rain pattern recognition and background recovery. Y. Yang et al. [31] proposed a segmentation aware progressive network via contrast learning, with three sub-networks for supervised rain removal, unsupervised background segmentation, and perceptual contrast, respectively. L. Cai et al. [32] extract depth and density information from rainy images, based on which conditional generative adversarial network is utilized to finish the job of rain removal. In [33], a novel attention-guided rain removal network was constructed, to simultaneously learn and model multiple rain streak layers under different phases. X. Cui et al. [34] develop a semi-supervised image deraining network with knowledge distillation (SSID-KD) for better exploiting real-world rainy images. C. -Y. Lin et al. [35] proposed a two-stage deep neural network to solve the rain removal problem, where the predicted rain streak component of the first stage became the input to the second stage to further localize possible rain pixels. Y. Wei et al. [36] proposed a novel generative adversarial network-based rain removal network, which used supervised and unsupervised processes in a unified form. K. Jiang *et al*. [37] proposed a novel multi-level memory compensation network for rain streak removal, in which the learning task was decomposed into multiple sub-tasks, and correspondingly several parallel sub-networks were built to solve these sub-problems separately. In [38], a knowledge distillation method, that captured long-range spatial and channel correlations to help the teacher network obtain more background details, was presented for image deraining.

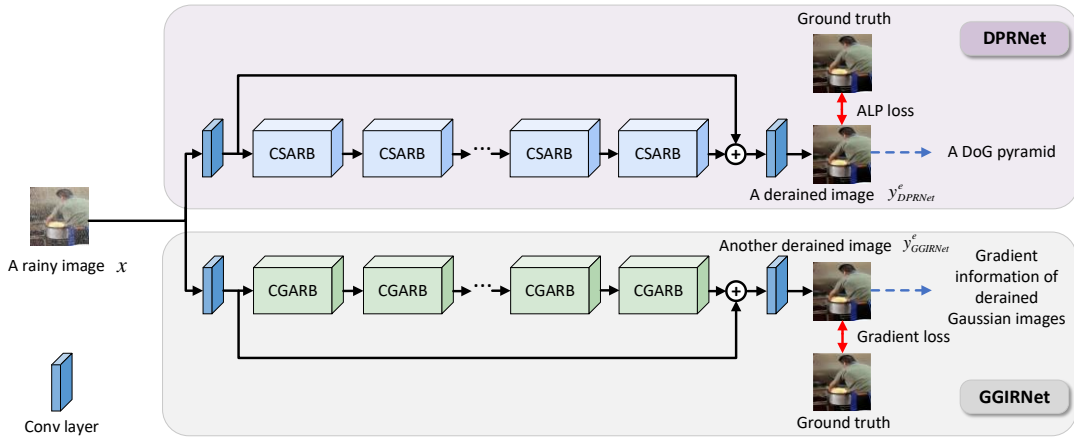

**Fig. 3.** Overall framework of our proposed IDSR.

Essentially, these aforementioned methods were always devoted to boosting the objective and subjective quality of an output derained image. However, different from existing human visual system (HVS) driven image deraining methods, in this work we completely concentrate on how to develop a task-driven approach to strengthen image feature supply for subsequent feature-based applications.

### III. PROPOSED IDSR

In this section, we will first describe the overall architecture of our proposed IDSR. Then, we put forward the DPRNet and GGIRNet, respectively. In the end, we give the detailed descriptions of the proposed ALP loss function.

#### A. Overview of IDSR

In this paper, an additive model usually adopted in existing methods is assumed where a rainy image $x$ is the superposition of its corresponding clean image (also called ground truth, GT) $y$ and a rain streak layer $r$, i.e.

$$x = y + r. \quad (1)$$

Then, after an image deraining network has learnt a rain streak estimation $r^e$, a derained image $y^e$ can be obtained as follows by subtracting $r^e$ from $x$:

$$y^e = x - r^e. \quad (2)$$

Fig. 3 illustrates the overall framework of the proposed IDSR. In this novel IDSR, both the DPRNet and GGIRNet take the same rainy image $x$ as their inputs, but learn two various rain streak estimations $r^e_{DPRNet}$ and $r^e_{GGIRNet}$, respectively. These two estimations are then subtracted from $x$ to build different derained images, i.e. $y^e_{DPRNet}$ and $y^e_{GGIRNet}$, respectively.

$$y^e_{DPRNet} = x - r^e_{DPRNet}, \quad (3)$$

$$y^e_{GGIRNet} = x - r^e_{GGIRNet}. \quad (4)$$

As shown in Fig. 3, the DPRNet mainly consists of several successive CSARBs, the proposed ALP loss, and a skip connection. But unlike the DPRNet, in the GGIRNet several consecutive CGARBs, a gradient-wise loss, and a skip connection are employed. With different losses and modules, these two networks work to realize two different purposes, respectively. One is recovering a derained DoG pyramid, and the other recovering the gradient information including magnitudes and orientations of derained Gaussian images.

Therefore, $y^e_{DPRNet}$ and $y^e_{GGIRNet}$ are utilized to compute their DoG pyramid and gradient Gaussian images, respectively, which are further leveraged to determine SIFT key point locations and descriptors, respectively. Finally, according to the obtained scale and spatial gradient space extrema locations and descriptors, SIFT key points of $y$ are recovered.

#### B. DPRNet

In the DPRNet, the input $x$ is first passed to a convolutional layer for shallow feature extraction, as depicted in Fig. 3. Then, we take several consecutive CSARBs to excavate deep features. Finally, the shallow features are added to these deep features, followed by a convolution, producing $y^e_{DPRNet}$.

Fig. 4 gives the schematic diagram of the CSARB. From this figure, one can observe that similar to the famous residual block (Resblock) in [39], a skip connection is adopted in the CSARB, to handle the exploding gradient and vanishing gradient problems. However, different from the Resblock, in addition to a convolutional layer and a ReLU activation used alternatively twice, we also accept two attention modules - a channel attention module (CAM) and a spatial attention module (SAM) to weight and grasp rain streak information. For more detailed information of the CAM and SAM, please refer to [40].

#### C. GGIRNet

In comparison with the DPRNet, our GGIRNet employs several consecutive CGARBs instead of CSARBs to capture deep features, as given in Fig. 3. Fig. 5 provides the diagram of the CGARB, from which one can see that it takes a similar structure to the CSARB but a different attention module.

Inspired by existing attention mechanisms, including the CAM and SAM, in this paper we propose a new attention mechanism, i.e. GAM. The role of the CAM is to assign a weight to each channel according to the information importance of different channels, making the network focus on the more useful ones. Unlike the CAM, the SAM is an adaptive spatial region selection mechanism providing each spatial region a weight. Because in the CGARB we commit to recover the gradients of derained Gaussian images, it is necessary to adaptively assign each region a weight on the basis of its gradient importance. Therefore, we specially design the proposed GAM for that purpose.

The schematic illustration of our proposed GAM is presented in Fig. 6. First, we apply a sobel convolution (SC) to calculate the gradient features $G_X \in \Re^{C \times H \times W}$ and $G_Y \in \Re^{C \times H \times W}$ of the GAM input $F_{GAM}^i$ in x-axis and y-axis directions, respectively, where $C$, $H$, and $W$ represent the channel number, the height and width of the feature, respectively. Then, $G_X$ and $G_Y$ are passed to two block groups for the generations of $G_X^{'}$ and $G_Y^{'}$, respectively, where each of these two groups is composed of a $1 \times 1$ convolutional layer especially to change the feature size from $C \times H \times W$ to $1 \times H \times W$, a $5 \times 5$ convolutional layer, a ReLU, a $5 \times 5$ convolutional layer, and a ReLU in sequential. Next, we concatenate $G_X^{'}$ and $G_Y^{'}$, followed by a convolution layer and a sigmoid activation, resulting in a gradient attention map $M_G$. Finally, we multiply $F_{GAM}^i$ by $M_G$ element-by-element to forge the GAM output $F_{GAM}^o$. The mechanism of the novel GAM can be expressed as:

$$G_X^{'} = ReLU(W_3 * (ReLU(W_2 * (W_1 * G_X + b_1) + b_2)) + b_3), \quad (5)$$

$$G_Y^{'} = ReLU(W_6 * (ReLU(W_5 * (W_4 * G_Y + b_4) + b_5)) + b_6), \quad (6)$$

$$M_G = \sigma(W_7 * ([G_X^{'}, G_Y^{'}]) + b_7), \quad (7)$$

$$F_{GAM}^o = F_{GAM}^i \otimes M_G, \quad (8)$$

where $[\cdot]$ represents the concatenate operation, $W_q, q=(1,2,\cdots 7)$ and $b_q, q=(1,2,\cdots 7)$ are the convolution matrices and bias vectors adopted in the GAM, respectively.

To make recovered gradients further closer to those of GT, in the GGIRNet we also take a loss function in gradient domain as follows.

$$L_{grad} = \sum_{j=1}^{5} \| \nabla_h y^{g,j} - \nabla_h y_{GGIRNet}^{e,g,j} \|_1 + \| \nabla_v y^{g,j} - \nabla_v y_{GGIRNet}^{e,g,j} \|_1, \quad (9)$$

where $y^{g,j}, (j=1,2,\cdots,5)$ and $y_{GGIRNet}^{e,g,j}, (j=1,2,\cdots,5)$ are the five Gaussian images of $y$ and $y_{GGIRNet}^e$, respectively, $\nabla_h$ and $\nabla_v$ are horizontal and vertical gradient operators, respectively, as well as $\| \cdot \|_1$ is the $\ell_1$ norm.

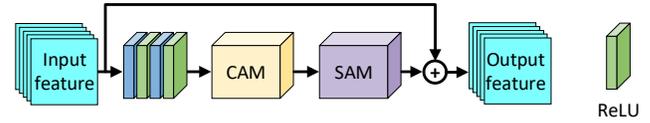

Fig. 4. Schematic diagram of the CSARB.

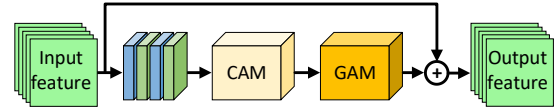

Fig. 5. Schematic diagram of the CGARB.

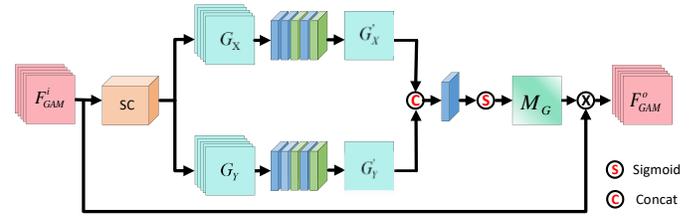

Fig. 6. Schematic illustration of our proposed GAM.

### D. Proposed ALP Loss

The ALP detector [22] is an efficient and powerful key point detector submitted to the 106th MPEG meeting and later adopted as an alternative scale space extrema detection solution in the MPEG CDVS (Compact Detector for Visual Search) standard. It is quite distinct from prior art, and has two clear advantages: one is that it is faster than most detectors, and the other is that its accuracy is also pretty good. In this paper, based on this excellent detector, we develop an ALP loss to try to detect as many key points as the corresponding clean image has for a derained image.

In this ALP solution, Laplacian of Gaussian (LoG) filtering is performed to sample scale space response (SSR) at pre-set scales, where LoG can detect the local extremum point and so in the SIFT method the DoG filter is introduced to approximate the LoG filter. The SSR is modelled as a polynomial function fitted with the following response:

$$h(m,n,\xi) \approx \sum_{k=1}^{K} \gamma_k(\xi) \cdot h(m,n,\xi_k), \quad (10)$$

where $h(m,n,\xi)$ is the LoG kernel at location $(m,n)$ and scale $\xi$, $\gamma_k(\xi)$ are functions of scale $\xi$, and $\xi_k (k=1,2,3,4)$ are four pre-defined scales.

Since $\gamma_k(\xi)$ is smooth and can be approximated by low-degree polynomials, the following third-degree polynomial of scale is used to represent $\gamma_k(\xi)$:

$$\gamma_k(\xi) \approx a_k \xi^3 + b_k \xi^2 + c_k \xi + d_k, \quad (11)$$

where $a_k$, $b_k$, $c_k$, and $d_k$ are coefficients.

In general, an image $I$ is directly convoluted with a LoG kernel to detect scale-invariant features and search extremums at multi-scale spaces as key points, which can be expressed as:

$$(I*h)[u,v,\xi] \approx \sum_{m=-w}^{w} \sum_{n=-w}^{w} I[u-m,v-n] \cdot \sum_{k=1}^{K} \gamma_k(\xi) \cdot h(m,n,\xi_k)$$
$$= \eta_3(u,v)\xi^3 + \eta_2(u,v)\xi^2 + \eta_1(u,v)\xi + \eta_0(u,v) \quad (12)$$

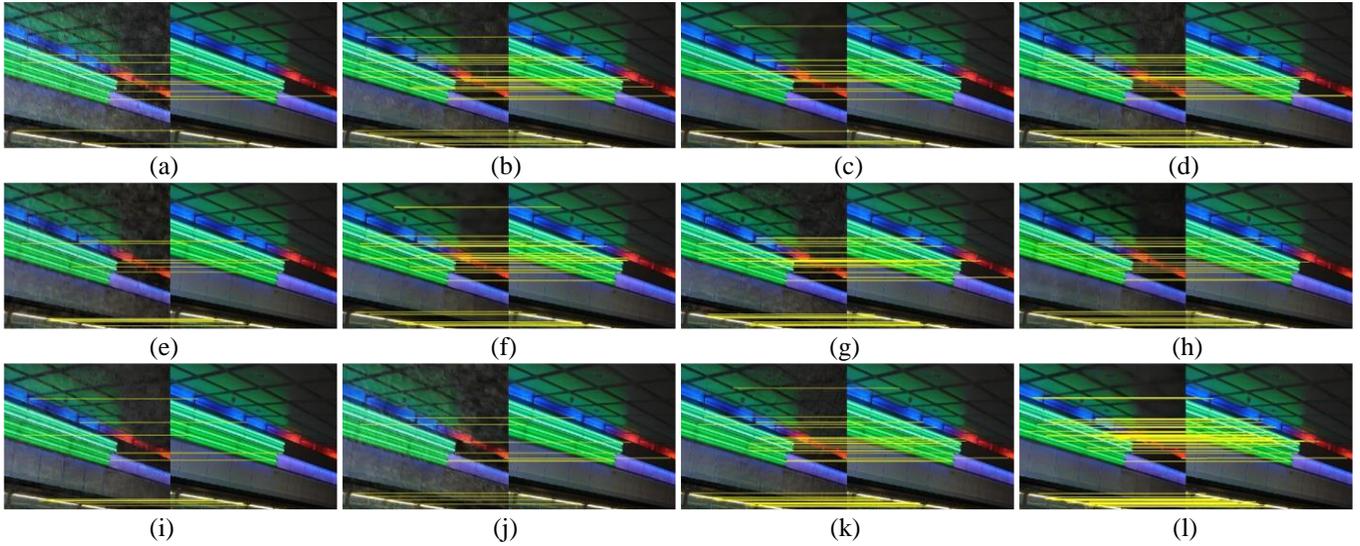

**Fig. 7.** Qualitative results on the first tested image pair selected from Rain1200 obtained by the: (a) DDN; (b) PReNet; (c) UMRL; (d) BRN; (e) ROMNet; (f) SSDRNet; (g) MPRNet; (h) MOSS; (i) ECNet; (j) SAPNet; (k) MAXIM; (l) Proposed, where the SIFT key points of a derained image (left) are matched with those of the corresponding clean image (right).

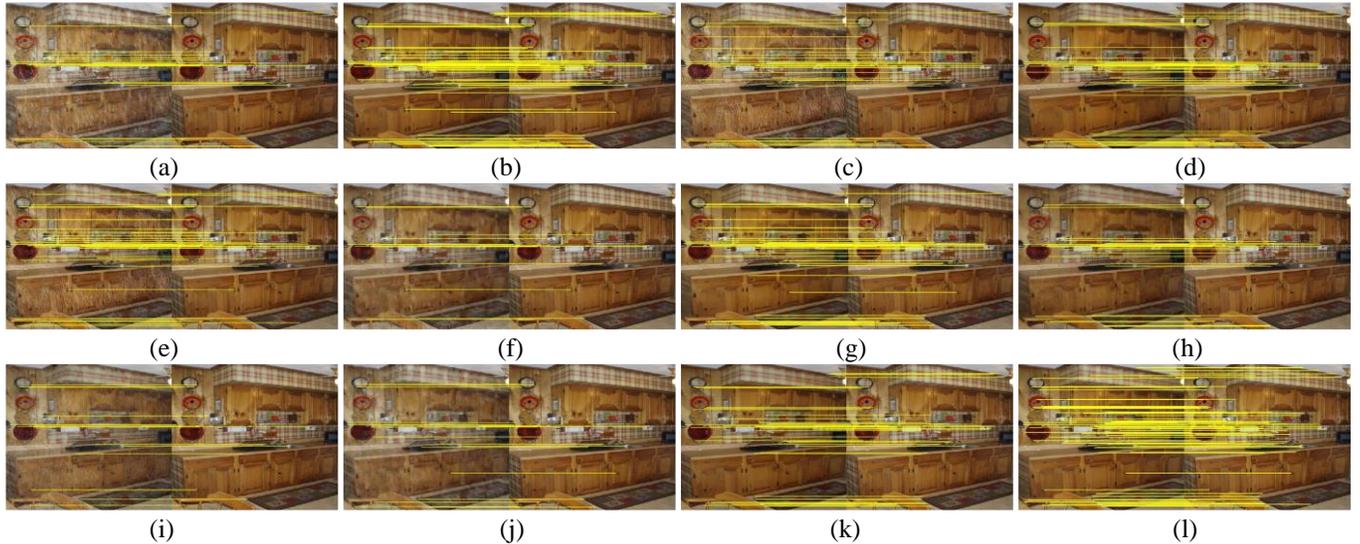

**Fig. 8.** Qualitative results on the second tested image pair selected from Rain1200 obtained by the: (a) DDN; (b) NLEDN; (c) PReNet; (d) UMRL; (e) BRN; (f) ROMNet; (g) SSDRNet; (h) MOSS; (i) ECNet; (j) SAPNet; (k) MAXIM; (l) Proposed, where the SIFT key points of a derained image (left) are matched with those of the corresponding clean image (right).

where $\eta_3(u,v)$, $\eta_2(u,v)$, $\eta_1(u,v)$ and $\eta_0(u,v)$ are the functions at $(u,v)$. We define this polynomial in (12) as the SSR at $(u,v)$.

Thus, in the ALP method key points are detected by finding the zeros in the derivative with respect to $\xi$ of SSR.

$$3\eta_3(u,v)\xi^2 + 2\eta_2(u,v)\xi + \eta_1(u,v) = 0 \Rightarrow \xi^*(u,v) = \xi. \quad (13)$$

It follows that, SSR is very important because it ultimately determines the locations of key points.

In the DPRNet, in order to more accurately locate SIFT key points, it is better to make sure that $y^e_{DPRNet}$ has the same key point locations with its corresponding clean version $y$.

Therefore, for the SSRs at $(u,v)$ of $y^e_{DPRNet}$ and $y$, their derivatives should be consistent as much as possible. According to (13), we can get

$$[3\eta_{y,3}(u,v) - 3\eta_{y^e_{DPRNet},3}(u,v)]\xi^2 + [2\eta_{y,2}(u,v) - 2\eta_{y^e_{DPRNet},2}(u,v)]\xi + [\eta_{y,1}(u,v) - \eta_{y^e_{DPRNet},1}(u,v)] \to 0, \quad (14)$$

where $\eta_{y^e_{DPRNet},j}(u,v), (j=1,2,3)$ and $\eta_{y,j}(u,v), (j=1,2,3)$ represent the functions $\eta_j(u,v), (j=1,2,3)$ used for $y^e_{DPRNet}$ and $y$, respectively.

To achieve (14), for each $\eta_j(u,v)$ its difference between these two images should be close to zero.

$$\eta_{y,1}(u,v) - \eta_{y^e_{DPRNet},1}(u,v) \to 0$$
$$\eta_{y,2}(u,v) - \eta_{y^e_{DPRNet},2}(u,v) \to 0 \quad (15)$$
$$\eta_{y,3}(u,v) - \eta_{y^e_{DPRNet},3}(u,v) \to 0,$$

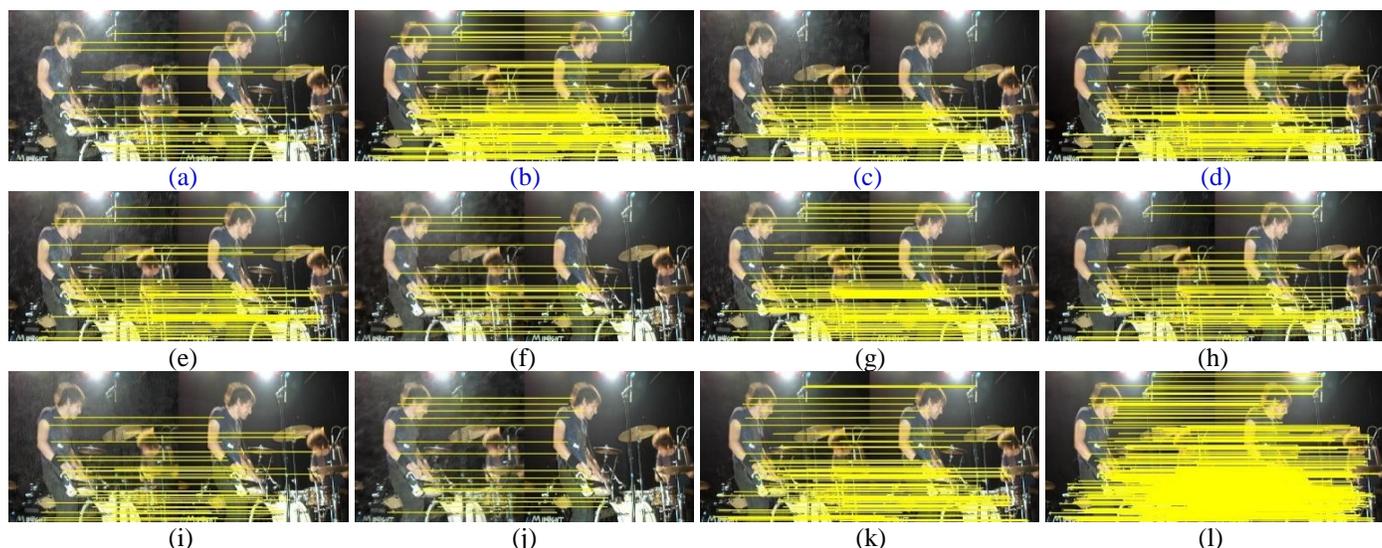

**Fig. 9.** Qualitative results on the third tested image pair selected from Rain1200 obtained by the: (a) DDN; (b) NLEDN; (c) PReNet; (d) UMRL; (e) BRN; (f) ROMNet; (g) SSDRNet; (h) MOSS; (i) ECNet; (j) SAPNet; (k) MAXIM; (l) Proposed, where the SIFT key points of a derained image (left) are matched with those of the corresponding clean image (right).

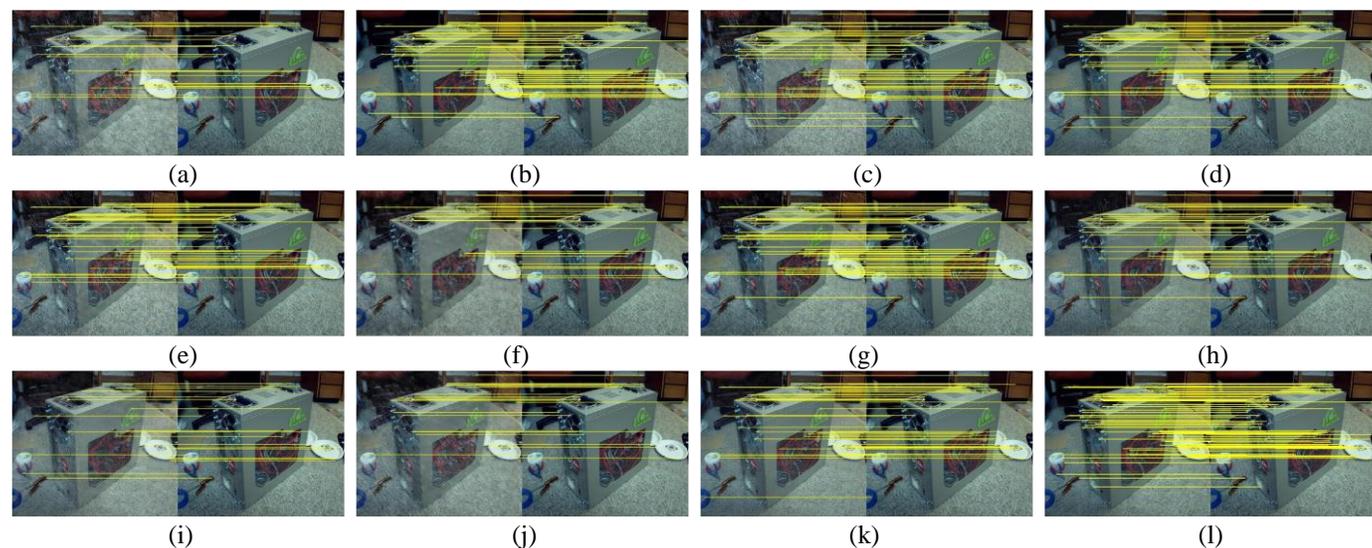

**Fig. 10.** Qualitative results on the fourth tested image pair selected from Rain1200 obtained by the: (a) DDN; (b) NLEDN; (c) PReNet; (d) UMRL; (e) BRN; (f) ROMNet; (g) MPRNet; (h) MOSS; (i) ECNet; (j) SAPNet; (k) MAXIM; (l) Proposed, where the SIFT key points of a derained image (left) are matched with those of the corresponding clean image (right).

Consequently, our proposed ALP loss is designed as follows:

$$L_{ALP} = \sum_{(u,v)} \sum_{j=1}^{3} |\eta_{y,j}(u,v) - \eta_{y^c_{DPRNet},3}(u,v)|. \qquad (16)$$

The adoption of this novel ALP loss function directly penalizes the distortion of the DoG pyramid in reconstructed grayscale images, and therefore has advantages over purely pixel L1 or L2 losses.

## IV. EXPERIMENTAL RESULTS

### A. Datasets and Implementation Details

To train and test our proposed IDSR method, we select three well-known and usually used public rainy image datasets, including Rain1200 [10], Rain1400 [9], and SPA-Data [41]. The former two synthetic datasets are mainly made up of rainy/clean image pairs in which each rainy image is generated by artificially adding rain streaks to its corresponding clean version. Three kinds of rain streaks: heavy, medium, and light, with their ratio of 1:1:1, are appended to 4000 clean images, respectively, producing the training dataset of Rain1200 containing 12,000 rainy/clean image pairs. On the other hand, Rain1200 also provides 1200 pairs of rainy/clean images for testing. Moreover, in Rain1400 there are 14,000 pairs of rainy/clean images, in which rainy images are synthesized from 1000 clean images plus rain streaks of different scales and orientations. In experiments, we choose 12,600 image pairs for training and the remaining 1400 image pairs for testing. Furthermore, since rain streaks in synthetic datasets may be quite different from those under real-world conditions, we also pick out the famous real-world rainy image dataset, i.e.

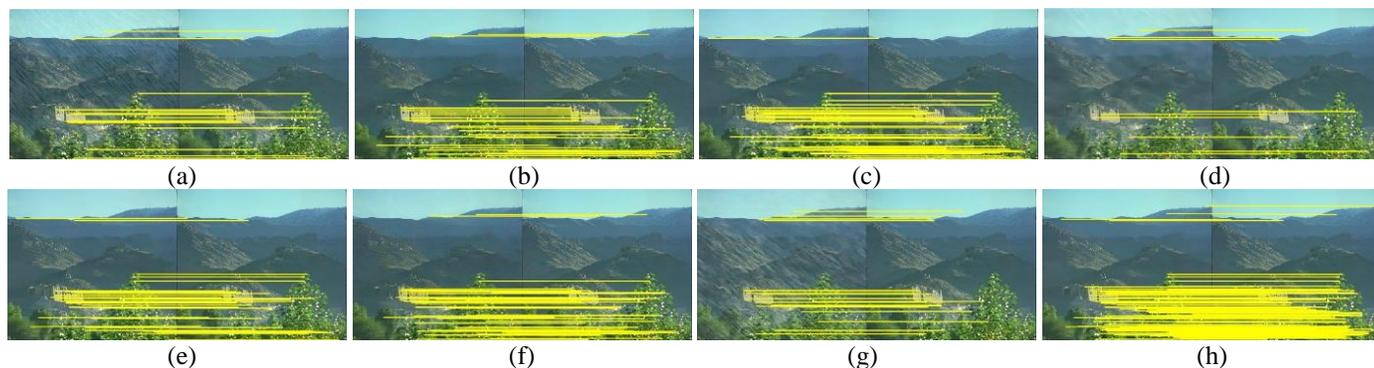

**Fig. 11.** Qualitative results on the first tested image pair selected from Rain1400 obtained by the: (a) DDN; (b) NLEDN; (c) BRN; (d) ROMNet; (e) SSDRNet; (f) MOSS; (g) SAPNet; (h) Proposed, where the SIFT key points of a derained image (left) are matched with those of its corresponding clean version (right).

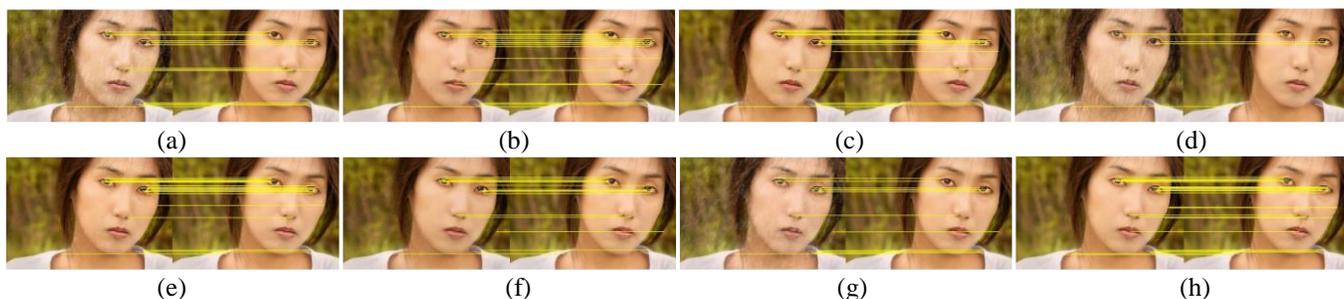

**Fig. 12.** Qualitative results on the second tested image pair selected from Rain1400 obtained by the: (a) DDN; (b) NLEDN; (c) BRN; (d) ROMNet; (e) PReNet; (f) MOSS; (g) SAPNet; (h) Proposed, where the SIFT key points of a derained image (left) are matched with those of its corresponding clean version (right).

SPA-Data, to test our proposed IDSR. In this SPA-Data, 638,492 rain/clean image pairs for training and 1000 image pairs for testing are included [42][43].

We implement our proposed IDSR in Python with PyTorch framework and train it with NVIDIA RTX 3080 GPU and 128 GB of memory on Ubuntu 20.4. Throughout the training of our proposed IDSR, the same settings are used for the DPRNet and GGIRNet. In the training process, the batch size and patch size are set to 16 and $128 \times 128$, respectively. We employ the Adam optimizer with an initial learning rate of 0.0001 which is then reduced by a factor of 0.5 every 20 epochs after 80 epochs in a total of 160 epochs. We use the derained image $y^e_{DPRNet}$ of the DPRNet to calculate PSNR and SSIM results.

### B. Baseline Methods

To verify the effectiveness of our proposed algorithm thirteen SOTA deep learning-based image deraining algorithms are compared, including DDN (deep detail network) [9], ECNet (embedding consistency and layered long short-term memory) [42], MOSS (memory oriented transfer learning for semi-supervised) [44], MPRNet (multi-stage progressive image restoration) [45], SAPNet (segmentation-aware progressive network) [31], ROMNet (Rain O'er Me) [46], SSDRNet (sequential dual attention network) [35], PReNet (progressive image deraining network) [12], BRN (bilateral recurrent network) [47], NLEDN (non-locally enhanced encoder-decoder network) [48], UMRL (uncertainty guided multi-scale residual learning) [49], MAXIM (multi-axis MLP)

[50], and CODE-Net (continuous density-guided network) [51].

In experiments we first should employ the available and runnable codes of each SOTA baseline approach to produce a derained image for every tested rainy image. Subsequently, the famous SIFT method is performed on each derained image to extract its SIFT key points. To generate derained Gaussian images, the scales of Gaussian images are set to 1.6000, 2.2627, 3.2000, 4.5255, and 6.4000, respectively.

### C. Qualitative Results Compared with SOTA Methods

Figs. 7, 8, 9, and 10 show the qualitative results obtained by the evaluated methods conducted on four chosen image pairs from Rain1200, respectively. To determine recovered SIFT key points, we match the key points extracted from every derained image with those of its corresponding clean version. Then, those matched SIFT key points are exactly what we are looking for. From Fig. 7, we can observe that some deraining methods, e.g. DDN and SAPNet, not only do not remove rain streaks well, but also regain few key points. Moreover, by comparing Figs. 7 (b) with 7 (c), it can be seen that PReNet obtains the derained image with more rain streak residues but recovers more SIFT key points than UMRL, implying that the quality of a derained image does not directly reflect its image feature recovery effect. This is the very reason that we develop a deraining algorithm especially for SIFT key point recovery from a single rainy image. Furthermore, from Figs. 8 (k) and 8 (l), one can find that MAXIM gets rid of rain streaks rather

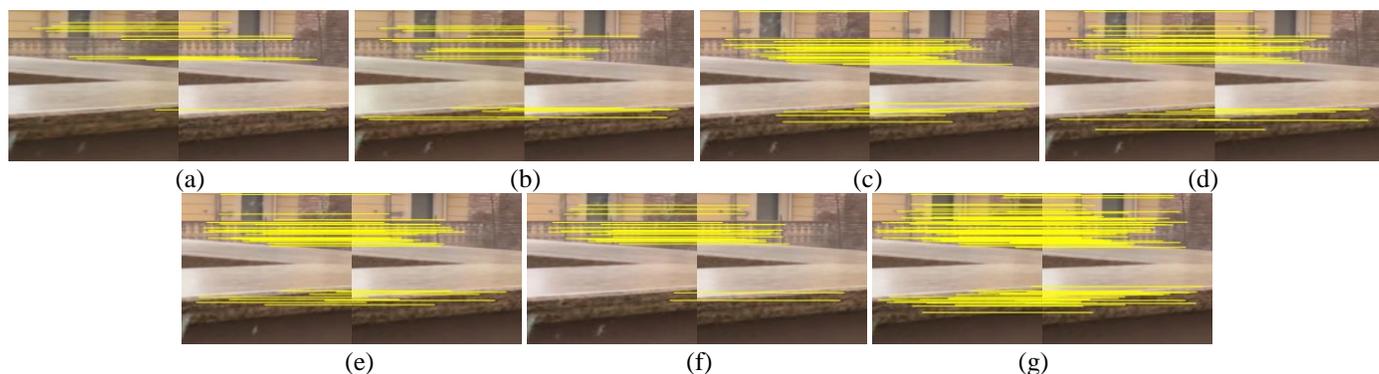

**Fig. 13.** Qualitative results on the first tested image pair selected from SPA-Data obtained by the: (a) PReNet; (b) BRN; (c) SSDRNet; (d) MPRNet; (e) CODE-Net; (f) SAPNet; (g) Proposed, where the SIFT key points of a derained image (left) are matched with those of its corresponding clean version (right).

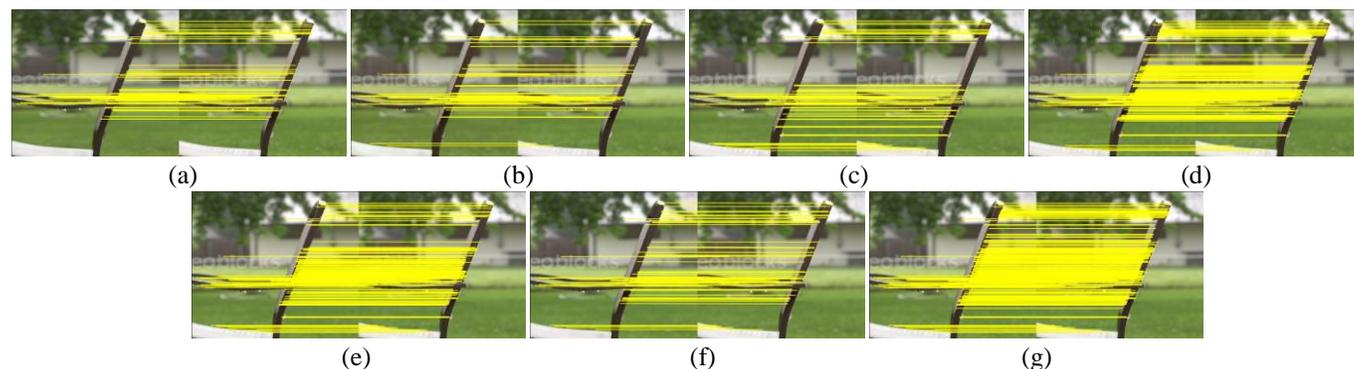

**Fig. 14.** Qualitative results on the second tested image pair selected from SPA-Data obtained by the: (a) PReNet; (b) BRN; (c) SSDRNet; (d) MPRNet; (e) CODE-Net; (f) SAPNet; (g) Proposed, where the SIFT key points of a derained image (left) are matched with those of its corresponding clean version (right).

well, and outputs some restored key points. However, in comparison with MAXIM, our proposed IDSR achieves an analogous rain streak elimination subjective satisfaction, whereas recovers much more key points. Similar results are also obtained from Figs. 9 as well as 10.

Figs. 11 and 12 also give the qualitative results via the evaluated methods conducted on two image pairs chosen from Rain1400, respectively. From Fig. 11, it can be observed that rain streaks still retain in the derained images built by both DDN and ROMNet. Comparatively, the latter method has a better performance of dislodging rain from an input rainy image, but produces fewer restored SIFT key points. Moreover, NLEDN not only eliminates more rain degraded components but also recovers more key points than ROMNet. Furthermore, our proposed IDSR establishes the derained image with a pretty good subjective quality and the most recovered key points among all the evaluated methods. Similar phenomena can be also found from Fig. 12.

The qualitative results of the two images selected from SPA-Data by the evaluation methods are given in Figs. 13 and 14, respectively. From Fig. 13, one can find that some rain stripes are not eliminated and still remain in each of the derained images obtained by PreNet, BRN, SSDRNet, MPRNet, CODE-Net, and SAPNet, resulting in insufficient recovered SIFT key points. Moreover, by comparing Figs. 13 (e) with 13 (g), although CODE-Net removes rain streaks better than other SOTA deraining methods, its recovered key points are still fewer than those of our IDSR. Furthermore, it can be seen from Fig. 14 that those derained images of the evaluated methods have similar subjective qualities. The proposed IDSR indeed improves the SIFT detection performance for the real rains.

According to the above qualitative results, there is a fact that it is perhaps that a derained image has a better quality but gets fewer recovered SIFT key points. Exploring its cause, the key one is luminance and structure losses are accepted in learning rain streak components to output a derained image with a high objective quality and a satisfied subjective quality, respectively. However, to recover more local image features, it is necessary to concentrate on the gradient difference between a derained image and its corresponding clean version, because gradient information plays a very important role in SIFT. As a result, our proposed IDSR exactly employs the novel ALP loss, the proposed GAM, and the gradient-wise loss to boost its key point recovery ability.

### D. Quantitative Results Compared with SOTA Methods

Table I presents the average quantitative evaluation results of the SOTA methods and our proposed IDSR on Rain1200. From the results in this table, we can see that both DDN and SAPNet restore relatively few SIFT key points of only about 83, as some degraded rain components are not removed yet by using these two methods. Moreover, NLEDN, SSDRNet, and MAXIM obtain the average recovered key points of more than 190, meaning that these three SOTA image deraining schemes perform rather well in restoring local image features. Furthermore, our proposed IDSR recovers average SIFT key points of 211.97, which not only is the largest among all the evaluated methods but also is much larger than our conference version [4], showing that the improvements over the method in [4] are highly necessary. In addition, compared with MPRNet, MAXIM yields a derained image with a lower PSNR result but more recoverd SIFT key points. From the results in this table, it can be seen that our proposed IDSR algorithm achieves the best average PSNR and SSIM values, while recovering the most SIFT key points.

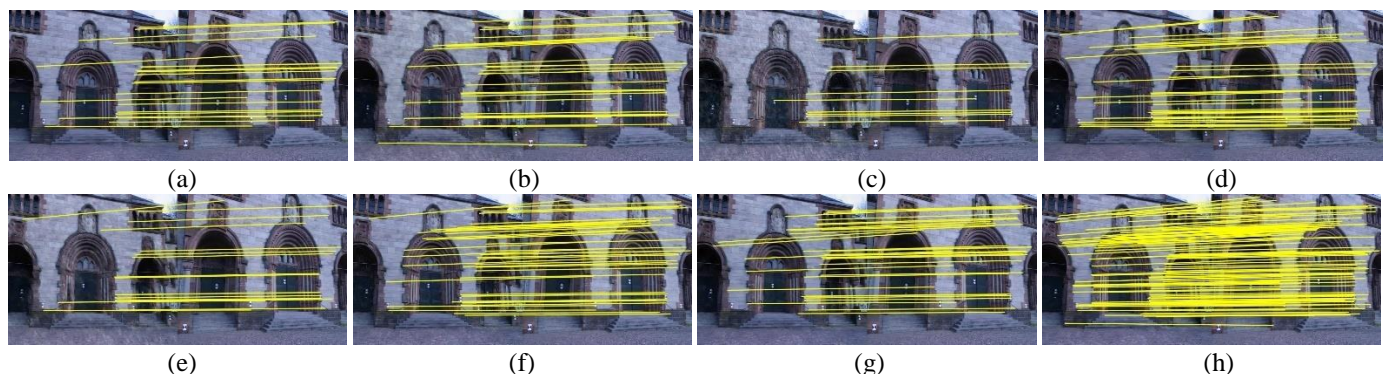

**Fig. 15.** Image matching results on the first tested image pair selected from DAISY obtained by the: (a) NLEDN; (b) BRN; (c) DDN; (d) MOSS; (e) PReNet; (f) SSDRNet; (g) UMRL; (h) Proposed, where the SIFT key points of a derained version (left) of one image are matched with those of the other image (right).

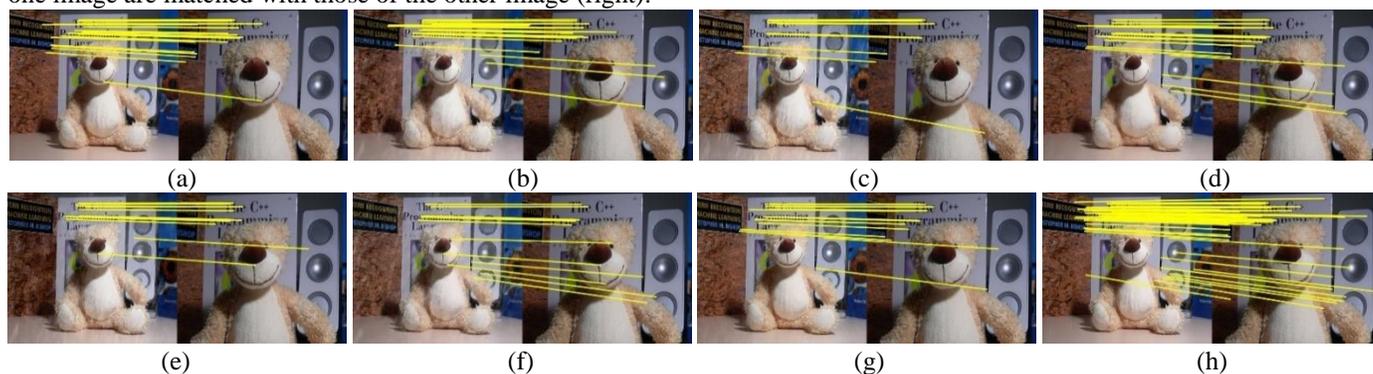

**Fig. 16.** Image matching results on the second tested image pair selected from DAISY obtained by the: (a) NLEDN; (b) BRN; (c) DDN; (d) MOSS; (e) PReNet; (f) SAPNet; (g) UMRL; (h) Proposed, where the SIFT key points of a derained version (left) of one image are matched with those of the other image (right).

The average recovered SIFT key point results of the SOTA methods and our proposed IDSR on Rain1400 are described in Table II. It should be noted that some evaluated methods have special requirements for tested image size, e.g. the multiples of 8 or 32. However, the Rain1400 dataset does not satisfy this size demand, so that we only use the methods without that image size limitation to compute recovery results, as illustrated in Table II. With the second best result as shown in Table I, NLEDN gets a good SIFT feature recovery effect on Rain1200 but merely takes fifth place on Rain1400. Moreover, although SSDRNet restores fewer average key points on Rain1200 than NLEDN, it ranks second on Rain1400. Furthermore, the number of average recovered key points on Rain1400 via our proposed IDSR stands at 225.31, the best recovery performance among all the evaluated approaches. Additionally, Table II also gives the average PSNR and SSIM results of these evaluated algorithms on Rain1400. One can see that although BRN gains higher average PSNR and SSIM values than SSDRNet, respectively, the recovered key points via the former are fewer than those of the latter. Therefore, it is necessary to design a task-driven image deraining method from the new perspective of image feature recovery. Compared with SOTA methods, our proposed IDSR not only recovers more key points but generates a derained image with better qualities.

To further validate the effectiveness of the proposed algorithm, we conducted experiments on the real-world dataset, i.e. SPA-Data, and compared our IDSR with SOTA algorithms. The results of the average recovered SIFT key points and average qualities are shown in Table III. From this table, it can be seen that by using the proposed IDSR, not only the average PSNR result, but also the value of average SSIM, are the best among all the methods. Moreover, the numbers of average key points by CODE-Net and SSDRNet are the second and third, respectively. Compared with these two methods, our proposed IDSR restores more average key points, where their gaps are 161.31 – 143.10 = 18.21 and 161.31 –

TABLE I
AVERAGE QUANTITATIVE RESULTS IN TERMS OF THE NUMBERS OF AVERAGE RECOVERED SIFT KEY POINTS AND PSNR/SSIM OF THE EVALUATED SOTA METHODS AND OUR PROPOSED ALGORITHM ON RAIN1200 DATASET, WHERE THE BEST, THE SECOND BEST, AND THE THIRD BEST RESULTS ARE MARKED WITH RED, BLUE, AND GREEN, RESPECTIVELY.

| Evaluated Methods | SIFT Points | PSNR/SSIM |
|---|---|---|
| DDN [9] (CVPR 2017) | 83.53 | 23.96/0.740 |
| NLEDN [48] (ACMMM 2018) | 199.01 | 33.32/0.923 |
| PReNet [12] (CVPR 2019) | 173.83 | 29.22/0.854 |
| UMRL [49] (CVPR 2019) | 183.49 | 30.83/0.901 |
| BRN [47] (TIP 2020) | 177.47 | 29.58/0.854 |
| ROMNet [46] (TIP 2020) | 132.03 | 29.10/0.879 |
| SSDRNet [35] (TIP 2020) | 194.73 | 32.98/0.919 |
| MPRNet [45] (CVPR 2021) | 185.28 | 31.57/0.898 |
| MOSS [44] (CVPR 2021) | 151.83 | 28.75/0.872 |
| ECNet [42] (WACV 2022) | 100.40 | 27.49/0.831 |
| SAPNet [31] (WACVW 2022) | 83.88 | 26.84/0.834 |
| MAXIM [50] (CVPR 2022) | 190.31 | 31.05/0.903 |
| Our preliminary [4] (VCIP 2021) | 150.55 | 30.37/0.901 |
| Proposed | 211.97 | 33.51/0.925 |

TABLE II
AVERAGE QUANTITATIVE RESULTS IN TERMS OF THE NUMBERS OF AVERAGE RECOVERED SIFT KEY POINTS AND PSNR/SSIM OF THE EVALUATED SOTA METHODS AND OUR PROPOSED ALGORITHM ON RAIN1400 DATASET, WHERE THE BEST, THE SECOND BEST, AND THE THIRD BEST RESULTS ARE MARKED WITH RED, BLUE, AND GREEN, RESPECTIVELY.

| Evaluated Methods | SIFT Points | PSNR/SSIM |
|---|---|---|
| DDN [9] (CVPR 2017) | 89.90 | 25.11/0.794 |
| NLEDN [48] (ACMMM 2018) | 195.90 | 30.82/0.911 |
| PReNet [12] (CVPR 2019) | 201.51 | 31.28/0.924 |
| BRN [47] (TIP 2020) | 207.88 | 31.43/0.926 |
| ROMNet [46] (TIP 2020) | 139.92 | 29.00/0.888 |
| SSDRNet [35] (TIP 2020) | 212.61 | 31.32/0.921 |
| MOSS [44] (CVPR 2021) | 172.79 | 28.30/0.893 |
| SAPNet [31] (WACVW 2022) | 100.02 | 27.11/0.856 |
| Proposed | 225.31 | 32.13/0.928 |

TABLE III
AVERAGE QUANTITATIVE RESULTS IN TERMS OF THE NUMBERS OF AVERAGE RECOVERED SIFT KEY POINTS AND PSNR/SSIM OF THE EVALUATED SOTA METHODS AND OUR PROPOSED ALGORITHM ON SPA-DATA, WHERE THE BEST, THE SECOND BEST, AND THE THIRD BEST RESULTS ARE MARKED WITH RED, BLUE, AND GREEN, RESPECTIVELY.

| Evaluated Methods | SIFT Points | PSNR/SSIM |
|---|---|---|
| PReNet [12] (CVPR2019) | 40.56 | 30.79/ 0.934 |
| BRN [47] (TIP 2020) | 46.87 | 31.70/ 0.939 |
| SSDRNet [35] (TIP 2020) | 116.16 | 36.66/ 0.963 |
| MPRNet [45] (CVPR 2021) | 101.54 | 33.19/ 0.945 |
| SAPNet [31] (WACVW 2022) | 55.26 | 32.93/ 0.943 |
| CODE-Net [51] (TMM 2023) | 143.10 | 39.32/ 0.979 |
| Proposed | 161.31 | 44.74/ 0.989 |

TABLE IV
AVERAGE QUANTITATIVE RESULTS IN TERMS OF THE NUMBERS OF AVERAGE RECOVERED SIFT KEY POINTS OF NLEDN, SSDRNET, AND OUR PROPOSED ALGORITHM ON THE TWO SYNTHETIC DATASETS, WHERE THE BEST AND THE SECOND BEST RESULTS ARE MARKED WITH RED AND BLUE, RESPECTIVELY.

| Evaluated Methods | RAIN1200 | RAIN1400 | Average |
|---|---|---|---|
| NLEDN [48] (ACMMM 2018) | 199.01 | 195.90 | 197.45 |
| SSDRNet [35] (TIP 2020) | 194.73 | 212.61 | 203.67 |
| Proposed | 211.97 | 225.31 | 218.64 |

116.16 = 45.15, respectively.

Finally, we especially compare our proposed IDSR with the second best methods on the two synthetic datasets, i.e. NLEDN and SSDRNet, respectively, and show related results in Table IV. From the results in this table, one can find that our proposed IDSR recovers the most SIFT key points on average for these two datasets, and the gap between the novel IDSR and NLEDN is up to 218.64 – 197.45 = 21.19.

*E. Ablation Studies*

In this section, we conduct the following ablation studies to demonstrate the effectiveness of each of our proposed methods.

*1) Ablation Study for the Proposed Divide-and-conquer Strategy*: In this subsection, we will verify the validity of our proposed divide-and-conquer strategy by removing the GGIRNet from the overall framework in Fig. 3. In contrast, we employ a one-task learning network by only training the DPRNet to create $y^e_{DPRNet}$. With the Gaussian images of the corresponding GT being the anchor, we can compute the mean square errors (MSEs) of the five derained Gaussian images via the one-task network and our proposed IDSR, respectively.

The average quantitative results of this ablation study conducted on Rain1200 are shown in Table V. From this table, we can see that each derained Gaussian image via our proposed IDSR has a lower MSE than that via the one-task method. Especially for the image with the scale of 1.6, the difference between these two schemes reaches the maximum. Experimental results show that our proposed IDSR generates more accurate derained Gaussian images compared with the one-task scheme, indicating that it is necessary to classify the SIFT recovery problem into two sub-problems and then employ the divide-and-conquer strategy to respectively solve them.

*2) Ablation Study for the Proposed ALP Loss*: To verify the superiority of our proposed ALP loss, we conduct an ablation study on this novel loss. In the study, we consider two different ways described as follows. First, we accept the widely used L2 loss to train the DPRNet instead of the ALP loss. Second, the proposed ALP loss is employed in the DPRNet as described in Section III. With the DoG pyramid of the corresponding GT being the anchor, we calculate the MSEs of the four derained DoG images obtained by the two methods, respectively.

Table VI presents the average quantitative results of this ablation study conducted on Rain1400. From this table, it can be seen that for each derained DoG image the proposed ALP loss gets a smaller MSE value than L2 loss, and thus considerably improves the accuracy of SIFT key point locations.

TABLE V
AVERAGE QUANTITATIVE RESULTS IN TERMS OF MSE OF ABLATION STUDY FOR THE PROPOSED DIVIDE-AND-CONQUER STRATEGY ON RAIN1200 DATASET, WHERE THE BEST IS MARKED WITH RED.

| Derained Gaussian images | One-task | Proposed |
|---|---|---|
| $Gau_{1.6000}$ | 9.2174 | 8.7313 |
| $Gau_{2.2627}$ | 7.2645 | 6.8562 |
| $Gau_{3.2000}$ | 5.5713 | 5.2370 |
| $Gau_{4.5255}$ | 4.1544 | 3.8822 |
| $Gau_{6.4000}$ | 3.0912 | 2.8581 |

TABLE VI
AVERAGE QUANTITATIVE RESULTS IN TERMS OF MSE OF ABLATION STUDY FOR THE PROPOSED ALP LOSS ON RAIN1400 DATASET, WHERE THE BEST IS MARKED WITH RED.

| Derained DoG images | L2 loss | Proposed |
|---|---|---|
| $DoG_{1.6000,2.2627}$ | 2.7088 | 2.5240 |
| $DoG_{2.2627,3.2000}$ | 2.3377 | 2.1601 |
| $DoG_{3.2000,4.5255}$ | 1.9810 | 1.8157 |
| $DoG_{4.5255,6.4000}$ | 1.7485 | 1.5955 |

TABLE VII
AVERAGE QUANTITATIVE RESULTS IN TERMS OF MSE OF ABLATION STUDY FOR THE PROPOSED GAM ON RAIN1400 DATASET, WHERE THE BEST IS MARKED WITH RED.

| Derained Gaussian images | w/o GAM | Proposed |
|---|---|---|
| $Gau_{1.6000}$ | 8.4561 | 8.3768 |
| $Gau_{2.2627}$ | 6.6430 | 6.5325 |
| $Gau_{3.2000}$ | 5.1625 | 4.9898 |
| $Gau_{4.5255}$ | 3.9333 | 3.7591 |
| $Gau_{6.4000}$ | 3.0928 | 2.8713 |

*3) Ablation Study for the Proposed GAM*: To demonstrate the performance of our proposed GAM, we conduct a corresponding ablation study for two situations of this module, which are described as follows. The first is that the proposed GAM is not adopted in the CGARB at all (denoted by w/o (without) GAM). In this case, after the CAM, the generated feature is directly added to the input feature to forge the output feature. The second is the complete use of the novel GAM in the CGARB, as illustrated in Fig. 5.

Table VII describes the average MSEs of the five derained Gaussian images for this ablation study conducted on Rain1400. From the table, it can be observed that the proposed algorithm produces those five images with their respectively smaller MSE values compared with the method without the GAM. Experimental results indicate that more precise derained Gaussian images are established by using the GAM, implying that this proposed GAM is of benefit to gradient extraction.

*F. Supplying for Image Matching*

In this section, we conduct extensive experiments to demonstrate the effectiveness of our proposed algorithm in strengthening image features supply for subsequent feature-based vision applications. Here we use an important computer vision task, i.e. image matching, as an example to illustrate the validity.

In experiments we select a well-known image matching dataset [20], i.e. the DAISY dataset [52], which consists of wide baseline image pairs with ground truth depth maps, including two short image sequences and some individual image pairs for evaluation. We use the rain synthesis method in Rain1200 [10] to add rain streaks to one image of each tested image pair. Afterward, we apply SOTA deraining methods and our proposed IDSR algorithm to remove rain and recover SIFT key points from every rainy image, respectively. Subsequently, SIFT matching is used to match the derained version of each rainy image with the other image of its corresponding image pair. Specifically, the SIFT features of every drained image are detected and described, and the putative correspondences are determined using the open source OpenCV library. The RANSAC algorithm [53] is then employed to remove misaligned points caused by incorrect matching.

Figs. 15 and 16 show the image matching results obtained by the evaluated methods conducted on two chosen image pairs from the DAISY dataset, respectively. From Fig. 15, it can be seen that our proposed IDSR obtains more matching SIFT key points than other SOTA methods, implying that our algorithm strengthens image feature supply for follow-up visual tasks. Similar phenomena can be also achieved from Fig. 16.

V. CONCLUSION

Different from existing HVS-driven image deraining approaches for pixel information recovery, in this paper we proposed an image deraining algorithm for SIFT key point recovery, dubbed IDSR. The proposed IDSR is a task-driven approach designed to bolster image feature supply for follow-up feature-based applications. Considering the essence of SIFT, we divide the recovery issue into two sub-problems, i.e. one being how to generate the DoG pyramid of a derained image, and the other being how to construct the gradients of derained Gaussian images. Consequently, we propose a divide-and-conquer strategy using two separate deep learning networks, including the DPRNet and GGIRNet, to solve these two sub-problems, respectively. Moreover, based on the notable ALP, an efficient and powerful key point detector adopted in the MPEG CDVS standard, in the DPRNet an ALP loss is advanced for the accurate SIFT extrema detection. In the ALP detector, the SSR at each pixel is modelled as a polynomial of scale, and key points are detected by finding the extreme points of all the polynomials. Thus, minimizing the ALP loss is designed to ensure the polynomials of an output derained image and those of its corresponding clean version remain consistent as much as possible. Furthermore, for the precise scale and spatial gradient space extrema description, in the GGIRNet we put forward a new attention mechanism, i.e. the GAM. This GAM is used to collect useful information to capture gradient-wise relationships. Finally, with the two different derained images generated by the DPRNet and GGIRNet, respectively, we compute their DoG pyramid and gradient information of Gaussian images, respectively, which are further applied to produce restored key points. Compared with SOTA methods in both quantitative and qualitative tests, experimental results demonstrate that our proposed scheme recovers more SIFT key points.

SIFT is one of the fundamental computer vision tools that has many important applications even at the era of deep learning. In this work, we demonstrated that a direct feature pyramid recovery via deep learning framework can be very effective in deraining for enhancing image feature supply for subsequent vision tasks like image matching. The proposed algorithm is robust and flexible, and can also be extended to other vision tasks.

REFERENCES

[1] Q. Wu, L. Wang, S. Huang, K. N. Ngan, H. Li, F. Meng, and L. Xu, "Subjective and objective de-raining quality assessment towards authentic rain image," *IEEE Trans. Circuits Syst. Video Technol.*, vol. 30, no. 11, pp. 3883-3897, Nov. 2020.
[2] K. Jiang, Z. Wang, P. Yi, C. Chen, Z. Han, T. Lu, B. Huang, and J. Jiang, "Decomposition makes better rain removal: an improved attention-guided deraining network," *IEEE Trans. Circuits Syst. Video Technol.*, vol. 31, no. 10, pp. 3981-3995, Oct. 2021.
[3] L. Zhu, Z. Deng, X. Hu, H. Xie, X. Xu, J. Qin and P.-A. Heng, "Learning gated non-local residual for single-image rain streak removal," *IEEE Trans. Circuits Syst. Video Technol.*, vol. 31, no. 6, pp. 2147-2159, Jun. 2021.
[4] P. Wang, W. Wu, Z. Li, and Y. Liu, "See SIFT in a rain: divide-and-conquer SIFT key point recovery from a single rainy image," in *Proc. IEEE Vis. Commun. Image Process.*, Munich, Germany, Dec. 2021, pp. 1-5.
[5] Y.-L. Chen and C.-T. Hsu, "A generalized low-rank appearance model for spatio-temporally correlated rain streaks," in *Proc. IEEE Int. Conf. Comput. Vis.*, Sydney, NSW, Australia, Dec. 2013, pp. 1968–1975.
[6] T. -X. Jiang, T. -Z. Huang, X. -L. Zhao, L. -J. Deng, and Y. Wang, "FastDeRain: A novel video rain streak removal method using directional gradient priors," *IEEE Trans. Image Process.*, vol. 28, no. 4, pp. 2089-2102, Apr. 2019.
[7] Y. Luo, Y. Xu, and H. Ji, "Removing rain from a single image via discriminative sparse coding," in *Proc. IEEE Int. Conf. Comput. Vis.*, Santiago, Chile, Dec. 2015, pp. 3397-3405.
[8] Y. Wang and T.-Z. Huang, "A tensor-based low-rank model for single-image rain streaks removal," *IEEE Access*, vol. 7, pp. 83437-83448, Jun. 2019.
[9] X. Fu, J. Huang, D. Zeng, Y. Huang, X. Ding, and J. Paisely, "Removing rain from single images via a deep detail network," in *Proc. IEEE Conf. Comput. Vis. Pattern Recognit.*, Honolulu, HI, USA, Jul. 2017, pp. 3855-3863.
[10] H. Zhang and V. M. Patel, "Density-aware single image de-raining using a multi-stream dense network," in *Proc. IEEE/CVF Conf. Comput. Vis. Pattern Recognit.*, Salt Lake City, UT, USA, Jun. 2018, pp. 695-704.
[11] X. Li, J. Wu, Z. Lin, H. Liu, and H. Zha, "Recurrent squeeze-and-excitation context aggregation net for single image deraining," in *Proc. Eur. Conf. Comput. Vis.*, Munich, Germany, Sep., 2018, pp. 262-277.
[12] D. Ren, W. Zuo, Q. Hu, P. Zhu, and D. Meng, "Progressive image deraining networks: a better and simpler baseline," in *Proc. IEEE Conf. Comput. Vis. Pattern Recognit.*, Long Beach, CA, USA, Jun. 2019, pp. 3937-3946.
[13] H. Zhang, V. Sindagi, and V. M. Patel, "Image de-raining using a conditional generative adversarial network," *IEEE Trans. Circuits Syst. Video Technol.*, vol. 30, no. 11, pp. 3943-3956, Nov. 2020.
[14] D. G. Lowe, "Distinctive image features from scale-invariant keypoints," *Int. J. Comput. Vis.*, vol. 60, no. 2, pp. 91-110, 2004.
[15] M. Ghahremani, Y. Liu, and B. Tiddeman, "FFD: fast feature detector," *IEEE Trans. Image Process.*, vol. 30, pp. 1153-1168, Dec. 2020.
[16] W. Zhou, L. Zhang, S. Gao, and X. Lou, "Gradient-based feature extraction from raw Bayer pattern images," *IEEE Trans. Image Process.*, vol. 30, pp. 5122-5137, May 2021.
[17] Y. Yao, Y. Zhang, Y. Wan, X. Liu, X. Yan, and J. Li, "Multi-modal remote sensing image matching considering co-occurrence filter," *IEEE Trans. Image Process.*, vol. 31, pp. 2584-2597, Mar. 2022.
[18] T. H. Phuoc and N. Guan, "A novel key-point detector based on sparse coding," *IEEE Trans. Image Process.*, vol. 29, pp. 747-756, Aug. 2020.
[19] A. Mustafa, H. Kim, and A. Hilton, "MSFD: multi-scale segmentation-based feature detection for wide-baseline scene reconstruction," *IEEE Trans. Image Process.*, vol. 28, no. 3, pp. 1118-1132, Mar. 2019.
[20] J. Ma, X. Jiang, A. Fan, and et al, "Image matching from handcrafted to deep features: a survey," *Int. J. Comput. Vis.*, vol. 129, pp. 23-79, 2020.
[21] F. Bellavia and C. Colombo, "Is there anything new to say about SIFT matching," *Int. J. Comput. Vis.*, vol. 128, no. 7, pp. 1847-1866, 2020.
[22] G. Francini, M. Balestri, and S. Lepsoy, "CDVS: Telecom Italia's response to CE1 – interest point detection," in *ISO/IEC JTC1/SC29/WG11, M31369*, Geneva, Switzerland, Oct. 2013.
[23] Y. Fu, L. Kang, C. Lin, and C. Hsu, "Single-frame-based rain removal via image decomposition," in *Proc. IEEE Int. Conf. Acoust. Speech Signal Process.*, Prague, Czech Republic, May 2011, pp. 1453-1456.
[24] C. -H. Son and X. -P. Zhang, "Rain removal via shrinkage of sparse codes and learned rain dictionary," in *Proc. IEEE Int. Conf. Multimedia Expo Workshops*, Seattle, WA, USA, Jul. 2016, pp. 1-6.
[25] D. -Y. Chen, C. -C. Chen, and L. -W. Kang, "Visual depth guided color image rain streaks removal using sparse coding," *IEEE Trans. Circuits Syst. Video Technol.*, vol. 24, no. 8, pp. 1430-1455, Aug. 2014.
[26] Y. Wang, S. Liu, C. Chen, and B. Zeng, "A hierarchical approach for rain or snow removing in a single color image," *IEEE Trans. Image Process.*, vol. 26, no. 8, pp. 3936-3950, Aug. 2017.
[27] Y. Ye, Y. Chang, H. Zhou, and L. Yan, "Closing the loop: joint rain generation and removal via disentangled image translation," in *Proc. IEEE/CVF Conf. Comput. Vis. Pattern Recognit.*, Nashville, TN, USA, Jun. 2021, pp. 2053-2062.
[28] Y. Yang and H. Lu, "A fast and efficient network for single image deraining," in *Proc. IEEE Int. Conf. Acoust. Speech Signal Process.*, Toronto, ON, Canada, Jun. 2021, pp. 2030-2034.
[29] Y. Yang, J. Guan, S. Huang, W. Wan, Y. Xu, and J. Liu, "End-to-end rain removal network based on progressive residual detail supplement," *IEEE Trans. Multimedia*, vol. 24, pp. 1622-1636, 2022.
[30] Y. Zheng, X. Yu, M. Liu, and S. Zhang, "Single-image deraining via recurrent residual multiscale networks," *IEEE Trans. Neural Netw. Learn. Syst.*, vol. 33, no. 3, pp. 1310-1323, Mar. 2022.
[31] S. Zheng, C. Lu, Y. Wu, and G. Gupta, "SAPNet: Segmentation-aware progressive network for perceptual contrastive deraining," in *Proc. IEEE/CVF Winter Conf. Appl. Comput. Vis. Workshops*, Waikoloa, HI, USA, Jan. 2022, pp. 52-62.
[32] L. Cai, Y. Fu, T. Zhu, Y. Xiang, Y. Zhang and H. Zeng, "Joint depth and density guided single image de-raining," *IEEE Trans. Circuits Syst. Video Technol.*, vol. 32, no. 7, pp. 4108-4121, Jul. 2022.
[33] K. Jiang, Z. Wang, P. Yi, C. Chen, Y. Yang, X. Tian, and J. Jiang, "Attention-guided deraining network via stage-wise learning," in *Proc. IEEE Int. Conf. Acoust. Speech Signal Process.*, Barcelona, Spain, May 2020, pp. 2618-2622.
[34] X. Cui, C. Wang, D. Ren, Y. Chen, and P. Zhu, "Semi-supervised image deraining using knowledge distillation," *IEEE Trans. Circuits Syst. Video Technol.*, early access, Jul. 2022.
[35] C.-Y. Lin, Z. Tao, A.-S. Xu, L.-W. Kang, and F. Akhyar, "Sequential dual attention network for rain streak removal in a single image," *IEEE Trans. Image Process.*, vol. 29, pp. 9250-9265, Sep. 2020.

[36] Y. Wei, Z. Zhang, Y. Wang, H. Zhang, M. Zhao, M. Xu, and M. Wang, "Semi-deraingan: a new semi-supervised single image deraining," in *Proc. IEEE Int. Conf. Multimedia Expo*, Shenzhen, China, Jul. 2021, pp. 1-6.
[37] K. Jiang, Z. Wang, P. Yi, C. Chen, X. Wang, J. Jiang, and Z. Xiong, "Multi-level memory compensation network for rain removal via divide-and-conquer strategy," *IEEE J. Sel. Topics Signal Process.*, vol. 15, no. 2, pp. 216-228, Feb. 2021.
[38] Z. Su, Y. Zhang, J. Shi, and X. -P. Zhang, "Recurrent network knowledge distillation for image rain removal," *IEEE Trans. Cogn. Dev. Syst.*, early access, 2021, doi: 10.1109/TCDS.2021.3131045.
[39] B. Lim, S. Son, H. Kim, S. Nah, and K. M. Lee, "Enhanced deep residual networks for single image super-resolution," in *Proc. IEEE Conf. Comput. Vis. Pattern Recognit. Workshops*, Honolulu, HI, USA, Jul. 2017, pp. 136-144.
[40] X. Lin, L. Ma, B. Sheng, Z. -J. Wang, and W. Chen, "Utilizing two-phase processing with FBLS for single image deraining," *IEEE Trans. Multimedia*, vol. 23, pp. 664-676, Apr. 2021.
[41] T. Wang, X. Yang, K. Xu, S. Chen, Q. Zhang, and R. W. H. Lau, "Spatial attentive single-image deraining with a high quality real rain dataset," in *Proc. IEEE Conf. Comput. Vis. Pattern Recognit.*, Jun. 2019, pp. 12262-12271.
[42] Y. Li, Y. Monno, and M. Okutomi, "Single image deraining network with rain embedding consistency and layered LSTM," in *Proc. IEEE/CVF Winter Conf. Appl. Comput. Vis.*, Waikoloa, HI, USA, Jan. 2022, pp. 3957-3966.
[43] H. Wang, Q. Xie, Q. Zhao, and D. Meng, "A model-driven deep neural network for single image rain removal," in *Proc. IEEE Conf. Comput. Vis. Pattern Recognit.*, Jun. 2020, pp. 3100-3109.
[44] H. Huang, A. Yu, and R. He, "Memory oriented transfer learning for semi-supervised image deraining," in *Proc. IEEE/CVF Conf. Comput. Vis. Pattern Recognit.*, Nashville, TN, USA, Jun. 2021, pp. 7728-7737.
[45] S. W. Zamir, A. Arora, S. Khan, M. Hayat, F. S. Khan, M.-H. Yang, and L. Shao, "Multi-stage progressive image restoration," in *Proc. IEEE/CVF Conf. Comput. Vis. Pattern Recognit.*, Nashville, TN, USA, Jun. 2021, pp. 14816-14826.
[46] H. Lin, Y. Li, X. Fu, X. Ding, Y. Huang, and J. Paisley, "Rain O'er Me: Synthesizing real rain to derain with data distillation," *IEEE Trans. Image Process.*, vol. 29, pp. 7668-7680, Jul. 2020.
[47] D. Ren, W. Shang, P. Zhu, Q. Hu, D. Meng, and W. Zuo, "Single image deraining using bilateral recurrent network," *IEEE Trans. Image Process.*, vol. 29, pp. 6852-6863, May 2020.
[48] G. Li, X. He, W. Zhang, and et al, "Non-locally enhanced encoder-decoder network for single image de-raining," in *Proc. ACM Multimedia Conf. Multimedia Conf.*, Oct. 2018, pp. 1056-1064.
[49] R. Yasarla and V. M. Patel, "Uncertainty guided multi-scale residual learning-using a cycle spinning CNN for single image de-raining," in *Proc. IEEE/CVF Conf. Comput. Vis. Pattern Recognit.*, Long Beach, CA, USA, Jun. 2019, pp. 8397-8406.
[50] Z. Tu, H. Talebi, H. Zhang, F. Yang, P. Milanfar, A. Bovik, and Y. Li, "MAXIM: Multi-axis MLP for image processing," in *Proc. IEEE Conf. Comput. Vis. Pattern Recognit.*, New Orleans, Louisiana, USA, Apr. 2022.
[51] L. Yu, B. Wang, J. He, G.-S. Xia, and W. Yang, "Single image deraining with continuous rain density estimation," *IEEE Trans. Multimedia*, vol. 25, pp. 443-456, 2023.
[52] E. Tola, V. Lepetit, AND P. Fua. "Daisy: An efficient dense descriptor applied to wide-baseline stereo," *IEEE Trans. Pattern Anal. Mach. Intell.*, vol. 32, no. 5, pp. 815-830, 2009.
[53] K. G. Derpanis. "Overview of the RANSAC algorithm," *Image Rochester NY*, vol. 4, no. 1, pp. 2-3, 2010.

**Wei Wu** (Member, IEEE) received the B.S., the M.S., and Ph.D. degrees from Xidian University, Xi'an, China, in 1998, 2001, and 2005, respectively. He is currently an associate professor with the School of Telecommunication Engineering, Xidian University. From 2007 to 2008 he was a postdoctoral researcher at Sejong University, Seoul, Korea. He holds 21 issued or pending patents, more than 30 publications in book chapters, journals, and conferences in his research field, which include image and video compression, video rate control, image processing, and understanding. He is the organizer of one special session of IEEE Visual Communication & Image Processing (VCIP) 2022.

**Hao Chang** received the B.S. degree in information engineering from Hainan University, Haikou, China, in 2020. He is currently pursuing the M.S. degree with the School of Telecommunications Engineering, Xidian University. His research interests include image deraining and image dehazing.

**Zhu Li** (Senior Member, IEEE) is a professor with the Dept of Computer Science & Electrical Engineering, University of Missouri, Kansas City, and the director of NSF I/UCRC Center for Big Learning (CBL) at UMKC. He received his PhD in Electrical & Computer Engineering from Northwestern University in 2004. He was a AFRL summer faculty at the UAV Research Center, US Air Force Academy (USAFA), 2016~2018, 2020~2023. He was a senior staff researcher with the Samsung Research America's Multimedia Standards Research Lab in Richardson, TX, 2012~2015, a senior staff researcher with FutureWei Technology's Media Lab in Bridgewater, NJ, 2010~2012, an assistant professor with the Dept of Computing, the Hong Kong Polytechnic University from 2008 to 2010, and a principal staff research engineer with the Multimedia Research Lab (MRL), Motorola Labs, from 2000 to 2008. His research interests include point cloud and light field compression, graph signal processing and deep learning in the next gen visual compression, image processing and understanding. He has 50+ issued or pending patents, 190+ publications in book chapters, journals, and conferences in these areas. He is an IEEE senior member, associate Editor-in-Chief for *IEEE Trans on Circuits & System for Video Tech,* associated editor for *IEEE Trans on Image Processing* (2020~), *IEEE Trans on Multimedia* (2015~2018), *IEEE Trans on Circuits & System for Video Technology* (2016~2019). He received the Best Paper Award at *IEEE Int'l Conf on Multimedia & Expo* (ICME), Toronto, 2006, and *IEEE Int'l Conf on Image Processing* (ICIP), San Antonio, 2007.